\documentclass[a4paper]{jpconf}
\usepackage{graphicx,amssymb}
\usepackage[square,sort&compress]{natbib}
\def\farcs{\hbox{\kern 0.13ex.\kern -0.95ex\raisebox{.9ex}{\scriptsize$\prime\prime$}\kern -0.1ex}}

\begin{document}
\title{The nuclear star cluster of the Milky Way}

\author{Rainer Sch\"odel$^{1}$, David Merritt$^{2}$ and Andreas Eckart$^{3}$}

\address{$^{1}$ Instituto de Astrof\'isica de Andaluc\'ia (CSIC),
  Apartado 3004, 18080 Granada, Spain\\
$^{2}$ Department of Physics and Center for Computational
             Relativity and Gravitation, Rochester Institute of Technology,
             Rochester, NY 14623, USA\\
$^{3}$ I.Physikalisches
  Institut, Universit\"at zu K\"oln, Z\"ulpicher Str.\ 77, 50937 K\"oln,
Germany}

\ead{rainer@iaa.es, merritt@astro.rit.edu, eckart@ph1.uni-koeln.de}

\begin{abstract}
The nuclear star cluster of the Milky Way is a unique target in the
Universe. Contrary to extragalactic nuclear star clusters, using
current technology it can be resolved into tens of thousands of
individual stars. This allows us to study in detail its spatial and
velocity structure as well as the different stellar populations that
make up the cluster. Moreover, the Milky Way is one of the very few
cases where we have firm evidence for the co-existence of a nuclear
star cluster with a central supermassive black hole, Sagittarius\,A*.
The number density of stars in the Galactic center nuclear star
cluster can be well described, at distances $\gtrsim1$\,pc from
Sagittarius\,A*, by a power-law of the form $\rho(r)\propto
r^{-\gamma}$ with an index of $\gamma\approx1.8$. In the central
parsec the index of the power-law becomes much flatter and decreases
to $\gamma\approx1.2$. We present proper motions for more than 6000
stars within 1\,pc in projection from the central black hole. The
cluster appears isotropic at projected distances $\gtrsim0.5$\,pc from
Sagittarius\,A*. Outside of 0.5\,pc and out to $1.0$\,pc the velocity
dispersion appears to stay constant. A robust result of our Jeans
modeling of the data is the required presence of
$0.5-2.0\times10^{6}\,M_{\odot}$ of extended (stellar) mass in the
central parsec of the Galaxy.
\end{abstract}

\section{Introduction}

Nuclear star clusters (NSCs) can be found at the photometric and
dynamical centers of the majority of galaxies. They are most easily
identified in galaxies with low surface brightness. Imaging at high
angular resolution, usually with the {\it Hubble Space Telescope}, has
been an indispensable tool in the discovery and study of these objects
\cite{Phillips1996AJ,Carollo1998AJ,Matthews1999AJ,Cote2006ApJS}. NSCs
have typically effective radii of a few parsecs, luminosities of the
order $10^{6}-10^{7}\,L_{\odot}$, and masses from a few times $10^{5}$
to a few times $10^{7}\, M_{\odot}$
\cite{Ferrarese2006ApJ,Walcher2006ApJ}. They are the densest known
star clusters in the Universe. It appears that similar scaling
relations exist between NSCs and the properties of their host galaxies
as between supermassive black holes and their host galaxies
\cite{WehnerHarris2006ApJ,Ferrarese2006ApJ,Balcells2007ApJ}. For an
overview of observational facts and theoretical ideas on NSCs, we
refer the reader to Thorsten B\"oker's review in this volume
\cite{Boeker2008AHAR}.

\begin{figure}[!h]
\centering \includegraphics[width=\textwidth]{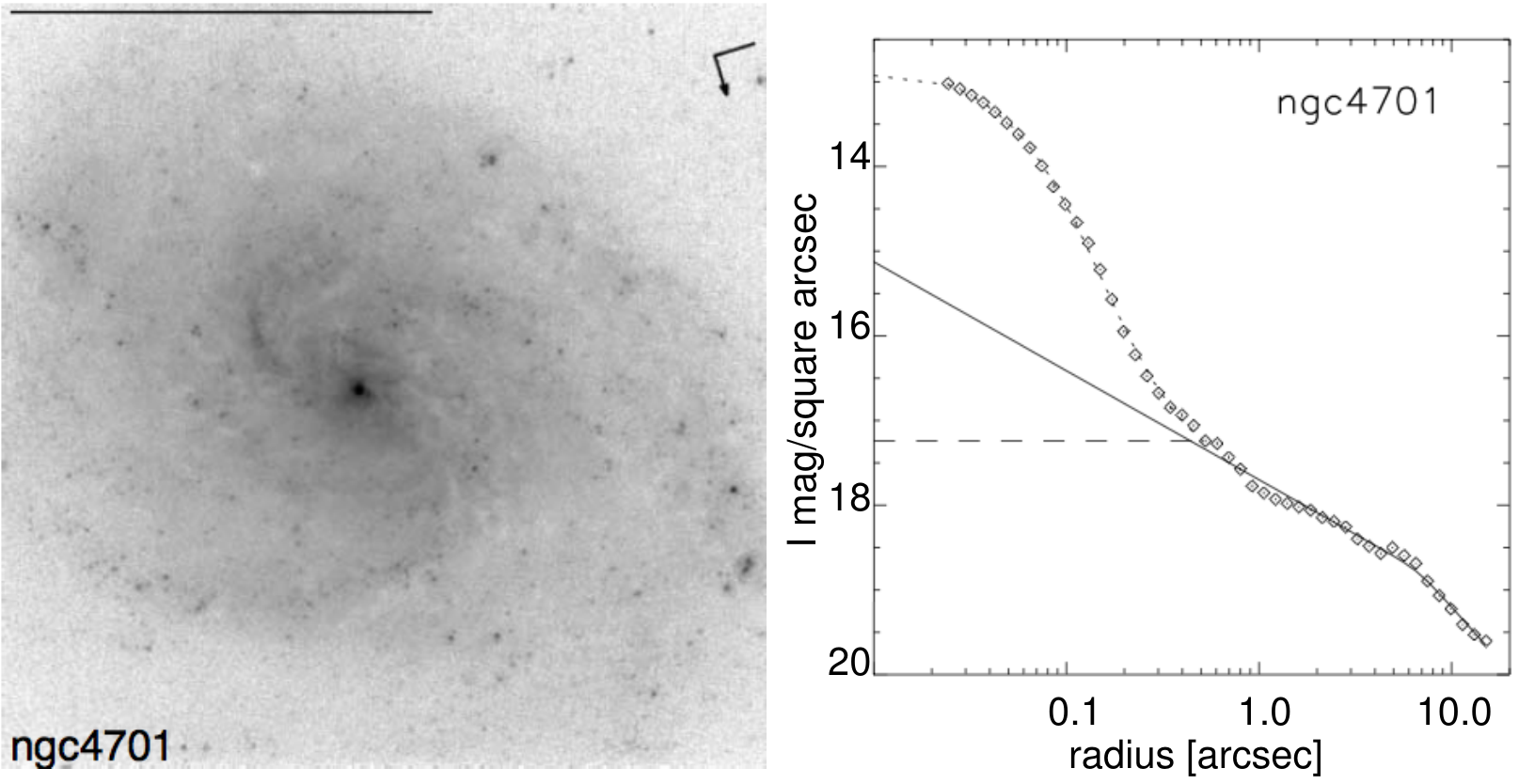}
\caption{\label{Fig:NGC4701} Example of a nuclear star cluster. Left:
  {\it HST} image of NGC~4701. The horizontal bar near the top of the
  image indicates a scale of 1\,kpc. The symbol in the upper right
  corner indicates north (arrow) and east directions. Right: $I$-band
  surface brightness profile of the central $10''$ of NGC~4701. Both
  images have been taken from \cite{Boeker2002AJ}$^{\dagger}$.\\
  $^{\dagger}${\footnotesize Reproduced by permission of the AAS.}}
\end{figure}

Figure\,\ref{Fig:NGC4701} shows an image of a typical NSC, along with
its light profile. It can be clearly seen how the NSC forms a highly
compact, separate entity that sticks out from its galactic
environment. An important problem in studying NSCs becomes apparent in
Figure\,\ref{Fig:NGC4701}: NSCs being extragalactic objects, it is hard
to study their properties in detail because they can be barely
resolved with current telescopes and instruments. This situation will
not change significantly with the advent of the next generation of
extremely large telescopes. Even with a resolution that is improved by
a factor of $5-10$, we will still be limited to studying the
integrated light of these objects. The interpretation of the data is
therefore complicated because NSCs are usually composed of stellar
populations from repeated star formation episodes
\cite{Walcher2006ApJ}. In many cases the most recent star formation
event took place not more than a few million years ago.

Studying the nuclear star cluster of the Milky Way offers the unique
advantage of being able to resolve it into individual stars. With a
distance of 8\,kpc
\cite{Reid1993ARA&A,Eisenhauer2005ApJ,Groenewegen2008A&A,Ghez2008ApJ}
to the Galactic center (GC), which will be assumed throughout this
work, one arcsecond corresponds to $\sim\,0.04$\,pc. The Milky Way
(MW) NSC can therefore be easily resolved.  Observations of the GC at
visible wavelengths are impossible ($A_{V}\approx 30$\,magnitudes,
e.g., \cite{Scoville2003ApJ}), therefore all observations have to be
carried out in the near- to mid-infrared.  In this wavelength regime
it is possible to reach scales of the order 1\,milli-pc, with
instruments such as the {\it HST}, but especially with adaptive optics
(AO) assisted imaging on ground-based 8-10\,m-class telescopes, like
the Keck or ESO VLT telescopes.

The NSC at the center of our Galaxy is highly interesting for a
further reason. Due to the observational difficulties involved in
analyzing external NSCs, there are only a few cases where there is
good evidence for the co-existence of an NSC with a central black
hole, e.g.\ in the case of NGC\,4395 \cite{Filippenko2003ApJ} and a
few other sources, e.g.\ \cite{Satyapal2007ApJ,Shields2008ApJ}. In
case of the Galactic center, however, stellar orbits have provided the
best evidence so far for the existence of a supermassive black hole,
e.g.\ \cite{Ghez2003ApJ,Schoedel2003ApJ,Eisenhauer2005ApJ}. The mass
of this black hole, coincident with the radio, X-ray, and infrared
source Sagittarius\,A* (Sgr\,A*) has been determined with high
precision to $4.0\pm0.2\times10^{6} M_{\odot}$ (for a fixed GC
distance of 8\,kpc)
\cite{Ghez2003ApJ,Eisenhauer2005ApJ,Ghez2005ApJ,Ghez2008ApJ}.

Although the GC has been a popular target of infrared observations
since the earliest moments when this technique became available
\cite{BecklinNeugebauer1968ApJ}, it has not been realized for a long
time that the MW contains an NSC. Although the NSC has been clearly
detected in the first observations
\cite{BecklinNeugebauer1968ApJ,Gatley1989IAUS}, it has probably not
been identified as such because of the difficulty in interpreting the
observations due to an unfortunate combination of circumstances:
low-angular resolution, perspective from within the Galactic disk,
strong and highly patchy extinction, and, above all, lack of the
knowledge of the existence of NSCs, which became only observable in
large numbers with the advent of the {\it HST}.  Since the beginning
of the 1990s, infrared (IR) observations of the GC were almost
exclusively directed toward the central (half-)parsec, and therefore
well within the NSC, in an effort to identify the central supermassive
black hole of the MW. The first authors to point out the existence and
derive the properties of the MW NSC were -- to my knowledge --
Launhardt et al.\ (2002) \cite{Launhardt2002A&A}.They derive an
effective radius of several parsecs, a near-infrared (NIR) luminosity
of $6\pm3\times10^{7}\,L_{\odot}$, and a mass of
$3\pm1.5\times10^{7}\,M_{\odot}$ for the MW NSC. Spectroscopic and
imaging studies have shown that there exist stellar populations of
different ages in the central parsec of the GC. The two most recent
star forming events have occurred about $10^{8}$ and a few times
$10^{6}$ years ago,
e.g.\ \cite{Allen1990MNRAS,Krabbe1995ApJ,Figer2004ApJ,Paumard2006ApJ,Maness2007ApJ}.

\begin{figure}[!h]
\centering
\includegraphics[width=\textwidth]{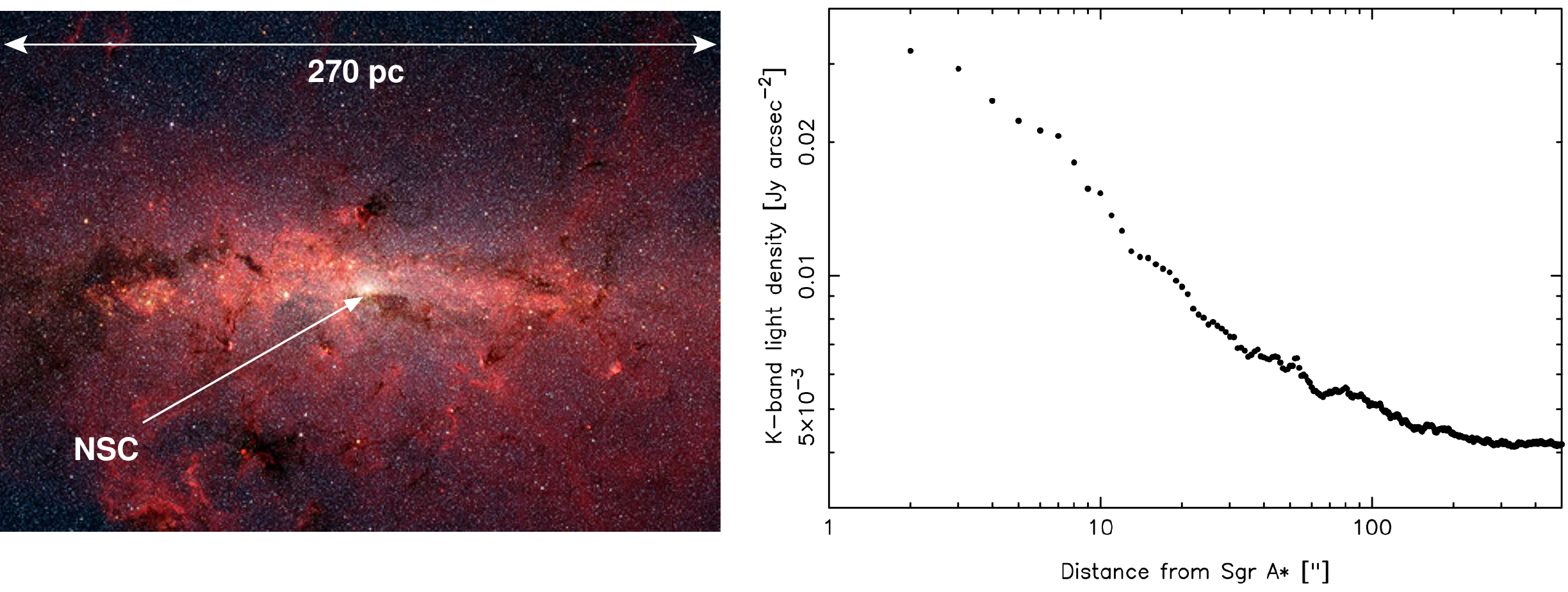}
\caption{\label{Fig:MW} The nuclear star cluster of the Milky
  Way. Left: Near-to-mid-infrared composite image from observations of
the {\it Spitzer Space Telescope} (Credit: NASA/JPL-Caltech/S. Stolovy
(Spitzer Science Center/Caltech)). Right: Light profile of the central
$500''$ (19\,pc at an assumed distance of 8\,kpc) of the Milky
Way. The profile was extracted from a 2MASS Ks-band image of the
GC. No extinction correction was applied. The flux calibration may be
subject to large systematic errors. The profile here is shown purely
for illustrative purposes.}
\end{figure}

The currently available observational evidence therefore characterizes
the MW NSC as a typical representative of its kind. An infrared
multi-wavelength false-color image of the GC environment, acquired
with the {\it Spitzer Space Telescope}, is shown in
Figure\,\ref{Fig:MW}. The NSC stands clearly out from its
environment. The right panel of Figure\,\ref{Fig:MW} shows the light
density vs.\ distance from Sgr\,A* as derived from 2MASS data
\cite{Skrutskie2006AJ}. It shows how the NSC sticks out from a rather
flat background.

In this contribution we will present recent findings on the
distribution and velocity dispersion of the  stars in the central
parsec of the Milky Way NSC.

\section{Density structure \label{sec:density}}

On large scales, the stellar density of the MW NSC can be well
described by a power-law, $\rho\propto r^{-\gamma}$, with
$\gamma\approx1.8$
\cite{BecklinNeugebauer1968ApJ,Catchpole1990MNRAS,Haller1996ApJ,Eckart1993ApJ}.
Early speckle imaging observations at the ESO 3.5\,m NTT indicated a
cluster core with a radius of about $5''$ or $0.2$\,pc
\cite{Eckart1993ApJ,Genzel1996ApJ}. Some indication for a central
density excess around Sgr\,A* was found in follow-up observations
\cite{Eckart1995ApJ}. However, ten years later, AO observations at the
ESO VLT showed unambiguously that the stellar density is not flat
around Sgr\,A*, but rises further inward, although with a
significantly smaller power-law index \cite{Genzel2003ApJ}. High
angular resolution was decisive for this discovery because it reduces
source confusion and thus the bias to underestimate the number of
detected sources. In the following, we will refer to this excess of
the stellar density above a flat core in the very center of the NSC as
the {\it cusp}. This cusp is, however, {\it not} to be confused with a
``classical'' cusp that may build up in a stellar cluster around a
black hole via two-body relaxation,
e.g.\ \cite{Bahcall1976ApJ,BahcallWolf1977ApJ,LightmanShapiro1977ApJ,Murphy1991ApJ,Merritt2008arXiv}.
First, classical cusp theory predicts a power-law index with
$-1.5\geq\gamma\geq-1.75$. This is significantly steeper than what has
been found by observations
\cite{Genzel2003ApJ,Schoedel2007A&A}. Second, it assumes largely
passive evolution of an undisturbed cluster with a single-age stellar
population. The latter is not the case in the MW NSC that has been
marked by repeated episodes of star formation and contains various
stellar populations (see introduction).

\begin{figure}[!h]
\centering
\includegraphics[width=\textwidth]{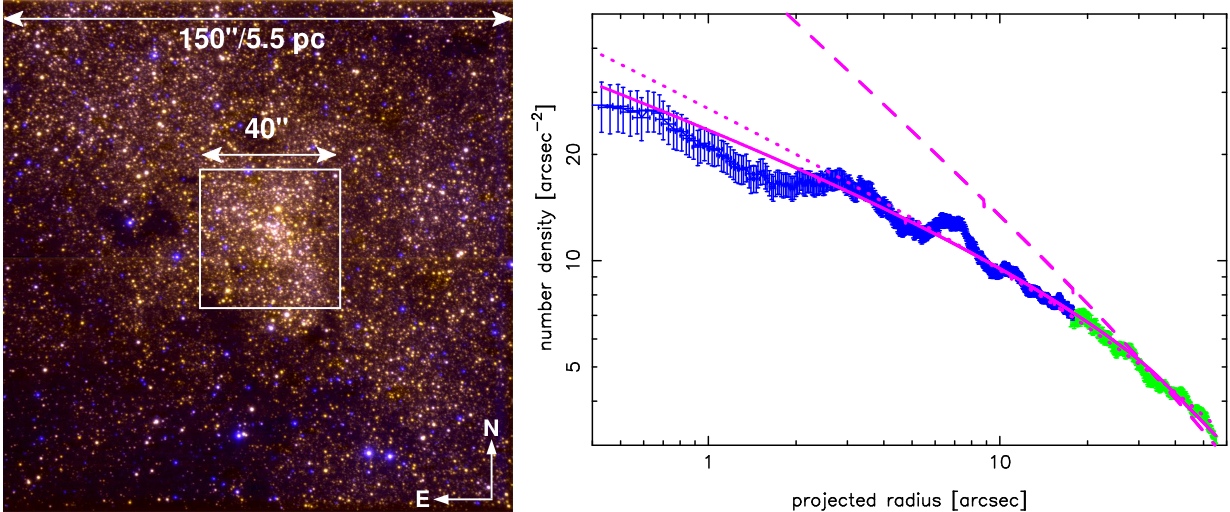}
\caption{\label{Fig:counts} Left: Seeing limited ISAAC/VLT false
  colour near-infrared ($Ks+J$) image of the GC used for stellar
  number counts \cite{Schoedel2007A&A}. The white rectangle indicates
  the area where AO observations were used to derive the stellar
  number density of the cluster. Right: Stellar number surface density
  vs.\ projected distance, $R$, from Sgr\,A*
  \cite{Schoedel2007A&A}. The blue data points and error bars are
  surface number density measurements extracted from adaptive optics
  observations (NaCo/VLT) and include stars down to
  mag$_{Ks}=17.5$. Crowding and extinction corrections were applied to
  the data. The green data points are (crowding and extinction
  corrected) surface number densities from seeing limited observations
  (ISAAC/VLT). They include only stars down to mag$_{Ks}=16.0$ and
  have been scaled to match the AO data in the overlap region. See
  \cite{Schoedel2007A&A} for details. The dotted line indicates the
  power-law slope that has been found for the outer parts of the MW
  NSC in many studies ($\gamma=1.8$).  The pink dashed line is a fit
  to the all data with a single power law ($\gamma = 1.45$). It fails
  to fit the data in the innermost arcseconds. The straight line is a
  fit with a broken power-law, with the power-law index in the
  innermost arcseconds being $\gamma=1.2$.}
\end{figure}

Further progress was achieved by an analysis that combined AO
observations (NaCo/VLT) with a large FOV of about $40''\times40''$ --
significantly beyond the cusp radius -- and seeing limited
observations (ISAAC/VLT) with a FOV of about $150''\times150''$, as
well as applied an improved data analysis technique
\cite{Schoedel2007A&A}. The left panel of Figure\,\ref{Fig:counts}
shows a $J$ + $K_s$-band false colour image of the corresponding
ISAAC/VLT observations. A plot of the derived (crowding and extinction
corrected) stellar surface number density vs.\ distance from Sgr\,A*
is shown in the right panel of Figure\,\ref{Fig:counts}. The stellar
surface density can be fit very well with a broken power-law. Inside a
projected break radius of $R_{\rm br}=6''\pm1''$, the power-law index
is $\Gamma_{\rm in}=0.2\pm0.05$, outside $R_{\rm br}$ the power-law
index is $\Gamma_{\rm out}=0.75\pm0.10$. After de-projection, these
values correspond to $\gamma_{\rm in}=1.2$ and $\gamma_{\rm
  out}=1.75$. The break radius cannot be constrained well after
de-projection. It ranges between $15''-30''$, but the assumption that
there is a clear-cut break radius is in any case an
over-simplification.  The value for $\gamma_{\rm out}$ is in excellent
agreement with earlier findings on the power-law index of the MW NSC
(see beginning of this section). The finding that the so-called {\it
  cusp} is extremely shallow explains why it has been mistaken as a
flat core in earlier, lower-resolution observations.

\begin{figure}[!b]
\centering
\includegraphics[width=\textwidth]{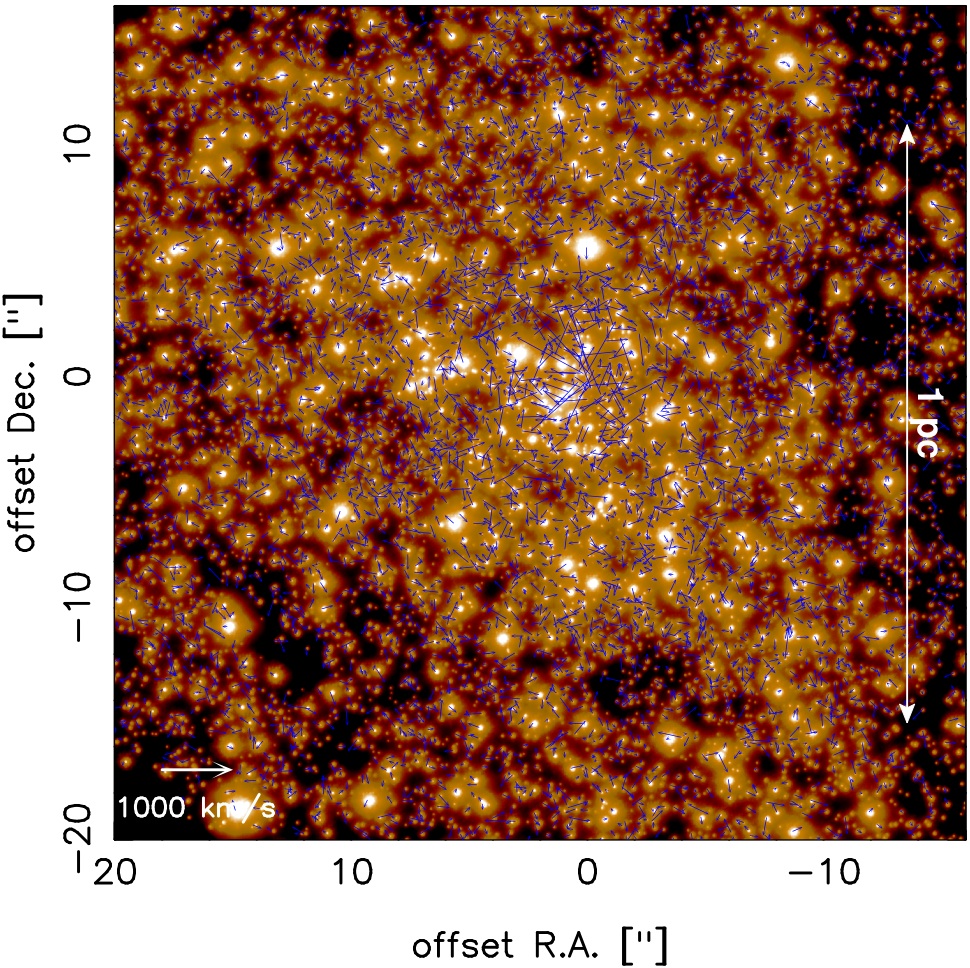}
\caption{\label{Fig:velmap} $K_s$-band map of the GC, derived from AO
  observations with ESO's NaCo/VLT (constructed from astrometric
  positions of the stars on 1 June 2006 and using a single point
  spread function for all stars). The arrows indicate the measured
  proper motions of the stars.  See Sch\"odel, Merritt \& Eckart (2008,
  submitted to A\&A) for details.}
\end{figure}
 
\section{Velocity structure of the MW NSC}

Several years of high-quality AO observations with a large FOV are now
available from a large number or programs that have used NaCo at the
ESO VLT to observe the GC. We have selected a suitable sub-set of
these observations (11 runs from spring 2002 to spring 2008) to
measure the proper motions of stars in the GC on a significantly
larger FOV than what has been published in previous work. Details on
the data processing are given in Sch\"odel, Merritt, \& Eckart (2008,
submitted to A\&A).

The proper motions of more than 6000 stars within 1\,pc of Sgr\,A*
were measured, with a mean uncertainty of just $12$\,km\,s$^{-1}$. The
measured velocities of the stars on the plane of the sky are
illustrated in the maps shown in Figure\,\ref{Fig:velmap}.

\begin{figure}[!h]
\centering
\includegraphics[width=\textwidth]{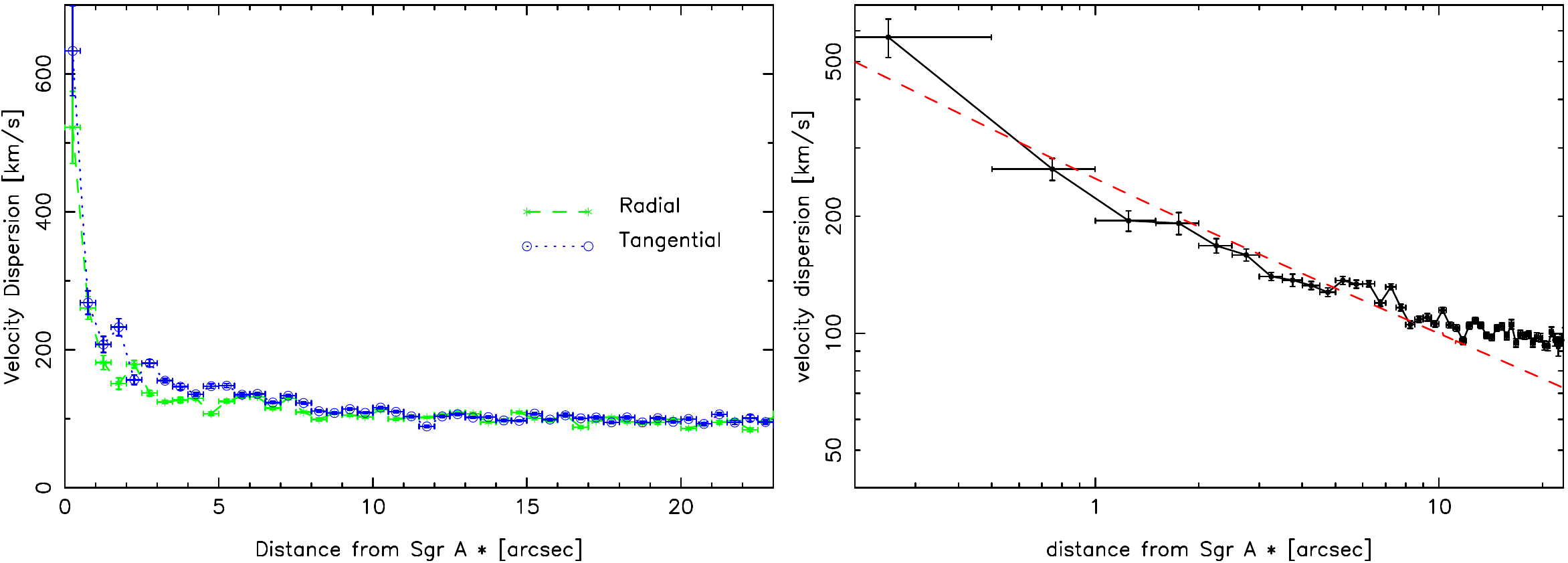}
\caption{\label{Fig:sigma} Left: Projected tangential (blue) and
  radial (green) proper motion velocity dispersion of stars with
  measured proper motions within 1\,pc of Sgr\,A*. A GC distance of
  8\,kpc was assumed. See Sch\"odel, Merritt \& Eckart (2008,
  submitted to A\&A) for details. Right: One-dimensional velocity
  dispersion of stars with measured proper motions within 1\,pc of
  Sgr\,A*.  It results from assuming isotropy and averaging the
  projected radial and tangential velocity dispersions. At $R>12''$
  the 1D velocity dispersion can be fit very well with a constant
  value.  The dashed red line indicates what a pure Kepler law would
  look like in projection, assuming a broken power-law structure of
  the stellar cluster with a break radius $R_{\rm break}=20.0''$ (the
  exact value has no significant influence on the result) and
  $\gamma=1.2$ inside and $\gamma=1.75$ outside of $R_{\rm break}$
  (see section\,\ref{sec:density}). The Kepler-law was fit to the data
  within $R=8''$. It can be seen that the velocity dispersion profile
  cannot be fit with a Kepler-law over the entire radial range of the
  data.  }
\end{figure}

The projected (with respect to Sgr\,A*) radial and tangential velocity
dispersion vs.\ distance from Sgr\,A* was computed from the proper
motions and is shown in the left panel of Figure\,\ref{Fig:sigma}. The
projected tangential velocity dispersion is somewhat higher than the
projected radial one in the region at $1''<R<6''$. Some net tangential
streaming motion is also present in this region (not shown here, but
see Sch\"odel, Merritt, \& Eckart, 2008, submitted to A\&A). These
findings are most probably related to the clockwise rotating disk of
massive, young stars that has been identified around Sgr\,A* (see
\cite{Genzel2003ApJ,Levin2003ApJ,Paumard2006ApJ,Lu2008arXiv} and
contributions by Hendrik Bartko, Jessica Lu, and Thibaut Paumard in
this proceedings). At $R\gtrsim8-10''$ the projected radial and
tangential velocity dispersions become equal within the measurement
uncertainties, providing evidence for an isotropic velocity structure
of the MW NSC. We find no signs of any overall rotation of the
NSC. Assuming isotropy, the one-dimensional velocity dispersion can be
derived. It is shown in the right hand panel of
Figure\,\ref{Fig:sigma}. A Keplerian fall-off of the velocity
dispersion as it would appear in projection on the sky (assuming
parameters for the cluster structure as discussed in
section\,\ref{sec:density}) is indicated in the plot. A pure
Kepler law would be expected if the gravitational potential were
dominated merely by the mass of the supermassive black hole
Sgr\,A*. The measured velocity dispersion deviates clearly from a
Kepler-law at distances $R\gtrsim12''$ ($R\approx0.5$\,pc). This
indicates that the potential of the distributed mass of the stellar
cluster must be taken into account when interpreting the measured
velocity dispersion. At $R\gtrsim12''$ ($\gtrsim\,0.5$\,pc) the
velocity dispersion can be fitted very well with a constant value of
$98.9\pm0.9$\,km\,s$^{-1}$.

Knowing both components of the proper motion velocity dispersion at
every radius permits a formally unique derivation of the mass profile
around Sgr\,A*, without the usual degeneracies associated with an
unknown anisotropy \cite{Leonard1989ApJ}. We applied these methods to
the proper motion data, as well as carrying out isotropic Jeans
modelling.  A robust result of our analysis was the presence of
$0.5-2.0\times10^{6}\,M_{\odot}$ of extended (probably stellar) mass
within 1\,pc of Sgr\,A*, in addition to the mass of the black
hole. The models cannot put strong constraints on the {\it
  distribution} of this mass, unfortunately. However, assuming that
the mass is distributed in the form of a power-law, $\rho(r)\propto
r^{-\alpha}$, the models allow for values $0\leq\alpha\leq1.5$. Values
as large as $\alpha=2.0$ appear to be safely ruled out.

\section{Summary}

The nuclear star cluster at the center of the Milky Way is the only
such object in the Universe in which a significant part of its
constituting stars can be observed individually. This situation will
not even change with the advent of telescopes of the 30-60\,m class in
the next decades. The NSC at the GC is also of great interest because
we know with certainty that the cluster at the GC co-exists with the
supermassive black hole Sgr\,A*.

The density of stars in the MW NSC rises like a power-law with an
index of $\gamma\approx1.8$ toward Sgr\,A*. This slope becomes
significantly flatter, however, in the central parsec, where it
decreases to a value of $\gamma\approx1.2$. Crowding is a serious
problem in observations of the MW NSC, even when AO instrumentation on
8-10\,m-class telescopes is applied. Current observations show a
surface density of $>20$ stars per square arcsecond within a projected
radius of $R=1''$ of Sgr\,A* (Figure\,\ref{Fig:counts}). This number
will certainly increase with further improvement of angular
resolution.

We have measured the proper motions of more than 6000 stars within
$R\approx1$\,pc of Sgr\,A*. These data supersede in quantity and
quality any existing studies of this kind. The new proper motion data
show that at $R\gtrsim0.5$\,pc the velocity dispersion ellipsoid is
close to isotropic and can be fit very well by a constant value of
$98.9\pm0.9$\,km\,s$^{-1}$. A Keplerian fall-off of the velocity
dispersion due to the point mass of Sgr\,A* can be observed
unambiguously only at $R\lesssim0.4$\,pc. Jeans models robustly
require the presence of $0.5-2.0\times10^{6}\,M_{\odot}$ at
$R\leq1$\,pc in addition to the mass of the central black hole.

\ack This publication makes use of data products from the Two Micron
All Sky Survey, which is a joint project of the University of
Massachusetts and the Infrared Processing and Analysis
Center/California Institute of Technology, funded by the National
Aeronautics and Space Administration and the National Science
Foundation. Rainer Sch\"odel is a Ram\'on y Cajal Post Doctoral
Research Fellow funded by the Spanish Ministry for Science and
Innovation and the Spanish Research Council (CSIC). DM was supported
by grants AST-0807910 (NSF) and NNX07AH15G (NASA).

\small
\bibliography{gc.bib}

\end{document}